\documentclass[times,twocolumn,authoryear]{revtex4-2}

\usepackage{framed,multirow}
\usepackage[pdftex]{graphicx}
\graphicspath{{figures/}}
\usepackage[T2A]{fontenc}
\usepackage[utf8]{inputenc}

\usepackage[russian,english]{babel}
\usepackage{amssymb}
\usepackage{latexsym}

\usepackage{amsmath}
\usepackage{cmap}
\usepackage{multirow}
\usepackage{layouts}

\usepackage{caption}
\usepackage{float}
\usepackage{placeins}

\usepackage[usenames]{color}

\usepackage{url}
\usepackage[dvipsnames]{xcolor}
\definecolor{newcolor}{rgb}{.8,.349,.1}

\usepackage[colorlinks=true,urlcolor=blue,citecolor=red,linkcolor=red,bookmarks=true]{hyperref}

\begin{document}

	\title{Electromagnetic emission from plasma with counter-streaming electron beams in the regime of oblique instability dominance}

	\author{V. V. Annenkov (В.В. Анненков)}
	\affiliation{Budker Institute of Nuclear Physics SB RAS, 630090, Novosibirsk, Russia}

	\author{E. P. Volchok (Е.П. Волчок)}
	\affiliation{Budker Institute of Nuclear Physics SB RAS, 630090, Novosibirsk, Russia}
	
	\author{I.V. Timofeev (И.В. Тимофеев)}

	\affiliation{Budker Institute of Nuclear Physics SB RAS, 630090, Novosibirsk, Russia}

	\begin{abstract}
	
		The problem of electromagnetic emission generation in plasma with electron beams is relevant both for practical applications and for interpretation  of radio emission processes in astrophysical systems. In this work, we consider the case of counter-propagating electron beams injection into plasma. Such systems may occur in cosmic plasma in the case of closely spaced particle acceleration regions, and they can also be implemented in laboratory facilities. Using particle-in-cell numerical simulations, we have shown that high beam-to-radiation conversion efficiency can be achieved in the case when beams excite small scale oblique plasma oscillations. In this case, the radiation is generated in the vicinity of the second harmonic of the plasma frequency. For such waves the surrounding plasma is transparent. It has been found that the efficiency and spectrum of the radiation are not dependent on the thickness of the beams. It has been also shown that parameters of the system necessary for efficient radiation generation by the discussed mechanism can be found using the exact linear theory of beam-plasma instability.
	\end{abstract}

\maketitle
\section{Introduction}

One of the fundamental problems of plasma physics is the interaction of plasma with fluxes of charged particles, in particular electron beams. Passing through plasma, electron beams cause the development of the two-stream (bump-on-tail) instability \cite{Akhiezer1949,Bohm1949}, which excites plasma oscillations. The efficiency of such oscillations excitation by electron beams can be quite high. While being a completely disrupted medium, the plasma can maintain electromagnetic (EM) oscillations of very high amplitude. The possibility of even a partial conversion of these oscillations into electromagnetic radiation may be promising for the creation of sources of powerful narrowband radiation in a wide range of frequencies \cite{Arzhannikov2020}. Also, electron beams have been considered in the context of heating and confinement of fusion plasma in open magnetic traps \cite{Burdakov2007,Burdakov2013} and as a promising way to create a target plasma \cite{Soldatkina2021}. Certain plasma conditions in tokamaks can lead to the formation of a beam of superthermal electrons (runaway electrons). Understanding of their interaction process with plasma in fusion devices can provide a useful diagnostic instrument and is important in maintaining the stability of the plasma \cite{Breizman2019,Hoppe2022}. Furthermore, as the current of beams increases in modern electron accelerators, the theoretical description of their interaction with residual gas and electromagnetic fields in  beamline becomes more of a plasma physics problem rather than vacuum electronics.  Another field in which the study of the beam-plasma interaction is extremely relevant is astrophysics \cite{Sanchez2019,Khotyaintsev2019,Lazar2023,Lee2022,Ziebell2021}. In various high-energy processes in stellar atmospheres, a significant number of charged particles are accelerated, and streams of ions and electrons capable of leaving the vicinity of stars are formed. We are particularly interested in beams originating from our nearest star, the Sun. Accelerated streams of charged particles generated during magnetic reconnection processes \cite{Pontin2022} or shock wave propagation \cite{Marcowith2016} lead to various physical processes in the solar atmosphere and can also leave it along open magnetic field lines, reaching Earth's orbit. Typical consequences of the propagation of such beams include plasma heating \cite{Lovelace1971,Cromwell1988}, generation of non-thermal radiation \cite{Ginzburg1958,Reid2014}, and various processes in natural magnetic traps \cite{Aschwanden2002}. In Earth's orbit, streams of charged particles can have a critical impact on the operation of electronic equipment \cite{Marusek2007}.

One of the fundamental processes in generating electromagnetic (EM) radiation for interpreting solar radio bursts is the three-wave coupling process of plasma waves $L$ and $L'$ into electromagnetic emission near the second harmonic of the plasma frequency $L+L'\rightarrow T_{2\omega_p}$ \cite{Ginzburg1958}. Typically, single-beam systems are considered, where the plasma waves traveling in the opposite direction  arise due to non-linear processes. However, a significant portion of the beam power is lost to plasma heating and excitation of non-radiating harmonics. High efficiency in radiation generation in single-beam systems can be achieved, for example, by introducing longitudinal density modulation in the plasma. This modulation allows the beam-driven modes  to convert their energy to superluminal plasma oscillations, which can resonantly excite vacuum electromagnetic modes at the plasma frequency \cite{Timofeev2015,Annenkov2016a,Glinskiy2022} and its second harmonic \cite{Annenkov2019c}. The required plasma structures can be self-consistently developed in initially homogeneous plasma under the influence of the modulation instability \cite{Annenkov2016b,Annenkov2019,Annenkov2023}, although this process requires some time and the presence of sufficient conditions for the development.

The increase in the level of EM emission from plasma with two counter-propagating beams compared to the case of a single beam is an experimental fact \cite{Leung1981,Intrator1984,Schumacher1993}. Such systems can be realized in the solar atmosphere in the presence of closely spaced particle acceleration regions. In laboratory plasma, it is possible to artificially implement such a system with the aim of creating a radiation source for practical applications. It has also recently been shown \cite{Annenkov2020} that in a two-beam system, it is possible to choose parameters (beam distribution, magnetic field, etc.) in such a way that the beam-excited plasma oscillations $L$ and $L'$ are capable of participating in a three-wave process $L+L'\rightarrow T_{2\omega_p}$ already during the linear stage. However, the approach proposed in \cite{Annenkov2020} has a significant drawback. For efficient emission, a fairly precise localization of the maximum growth rate of beam instability in the region of the three-wave process is required. As simulations have shown, even relatively small deviations of system parameters from the found efficient regime lead to a significant decrease in the emission level. Therefore, the first problem lies in the need for fine-tuning the system to achieve the efficient regime, while the second problem is that various natural factors such as inhomogeneity of plasma density or magnetic field can easily take the system out of this regime.

In the discussed mechanisms, radiation is generated by plasma oscillations, while electron beams act merely as drivers that excite these oscillations.  However, plasma waves can also be driven by short laser pulses. The possibility of using laser beams in plasma to generate THz radiation is being actively investigated \cite{Kalmykov2020,Ashish2023,Lee2023}. One of a highly efficient mechanism  \cite{Timofeev2017b,Volchok2021} for radiation generation at doubled the plasma frequency occurs when plasma waves with different transverse structures and characteristic transverse sizes of the order of $c/\omega_p$, where $c$ is the speed of light in vacuum and $\omega_p$ is the plasma frequency, collide in the plasma. The specific excitation method of the plasma oscillations is not crucial for this mechanism to work. It has been demonstrated that this process also occurs when thin counter-propagating electron beams with different transverse sizes are injected into the plasma \cite{Annenkov2018,Annenkov2021,Kumar2022}.

In real systems, electron beams have significantly larger transverse sizes than the scale of $c/\omega_p$. Furthermore, the excitation of purely longitudinal plasma oscillations is not necessarily a characteristic feature. The work \cite{Annenkov2019a} demonstrates the fundamental possibility of this mechanism operating in plasma with wide counter-propagating relativistic beams, which excite oblique plasma oscillations with a transverse wavelength on the scale of $c/\omega_p$. The purpose of this paper is to investigate such a regime of the beam-plasma interaction and the generated electromagnetic radiation in more detail. Section \ref{sec:rad} gives basic information about the mechanism of EM emission generation at the second harmonic of plasma frequency due to interaction of counter plasma oscillations with different transverse structures.  In section \ref{sec:param}, the parameters of the beam-plasma system, which will be further investigated by means of numerical simulations using the particle-in-cell method are described. The choice of the parameters is done using an exact kinetic linear theory of the beam-plasma instability, taking into account  the influence of an external magnetic field and relativistic effects.  Section \ref{sec:setup} describes the numerical schemes and the computational domain that were used. Section \ref{sec:result} contains the simulation results. The subsection \ref{subsec:rel} discusses the injection into plasma of beams with the average velocity $v_b=0.9c$, and the subsection \ref{subsec:subrel} -- with the average velocity $v_b=0.7c$. In both cases, the injection of a single beam into plasma is initially examined to investigate the convergence in the number of model macro-particles and determine the length at which the beam relaxes. Then the injection of counter beams is investigated and the dependence of the generated radiation on their transverse size is studied. In subsection \ref{subsec:diffbeams}, it is shown that the discussed mechanism also works in the case of counter beams with different average speeds and the radiation is generated at an angle to the system axis. In section \ref{sub:disc_conc}, a discussion of the obtained results and suggestions for their further application are given.

\section{Radiation mechanism}\label{sec:rad}
\begin{figure*}
	\begin{center}
		\begin{minipage}[h]{0.49\linewidth}
			
			\includegraphics[width=1\linewidth]{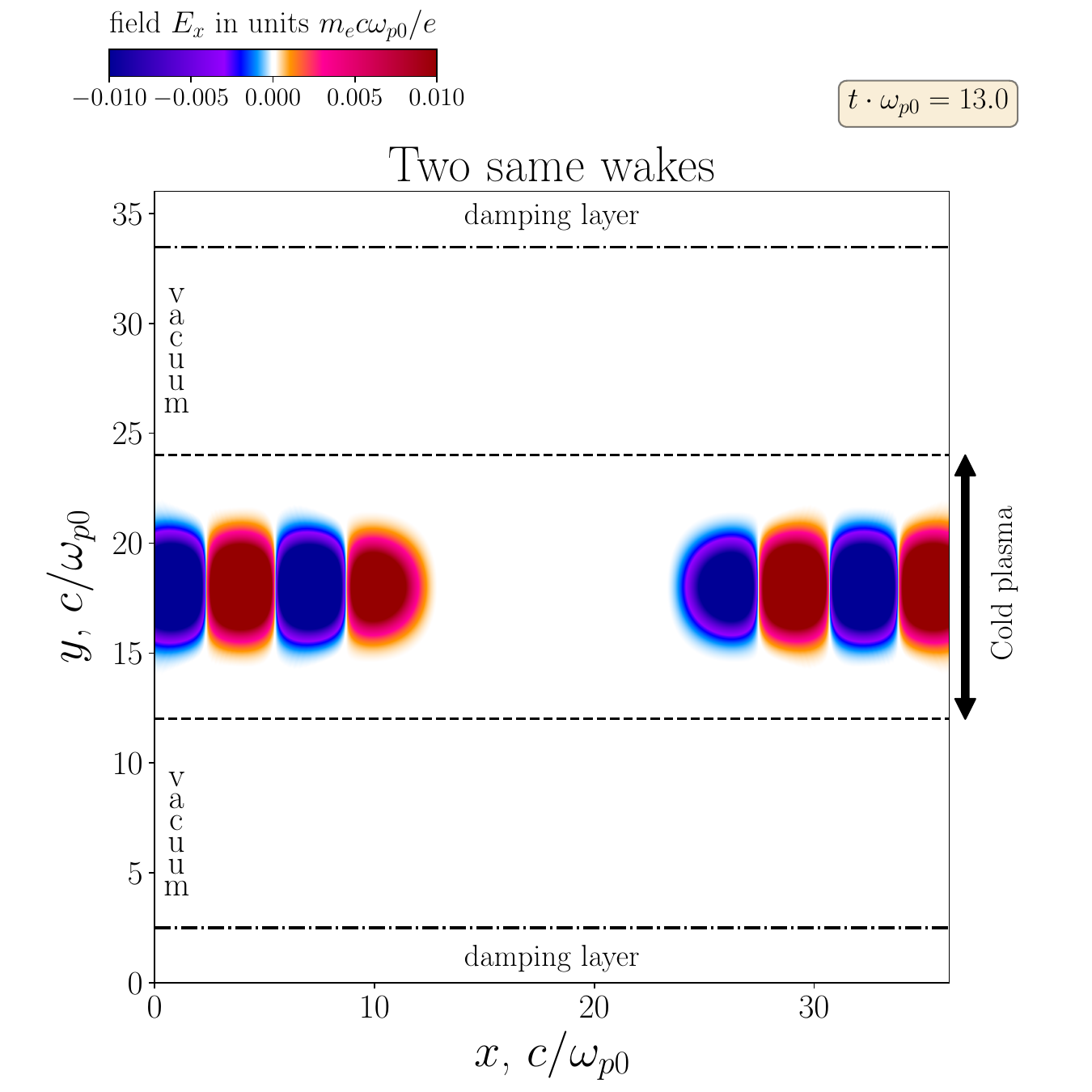}
			\includegraphics[width=1\linewidth]{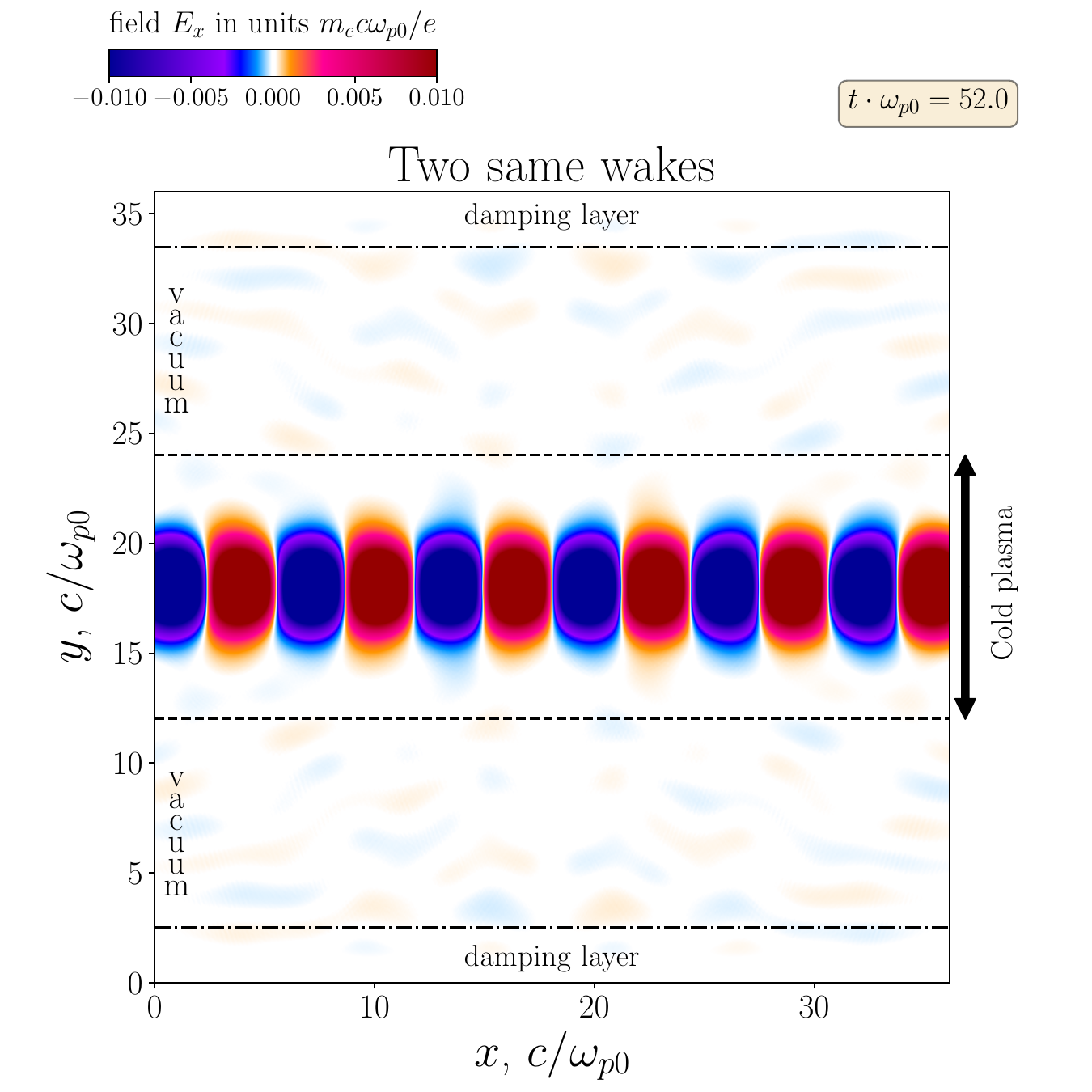}
			\caption{Counter plasma waves with the same transverse shape.}\label{fig:sameWakes}
		\end{minipage}
		\hfill
		\begin{minipage}[h]{0.49\linewidth}
			
			\includegraphics[width=1\linewidth]{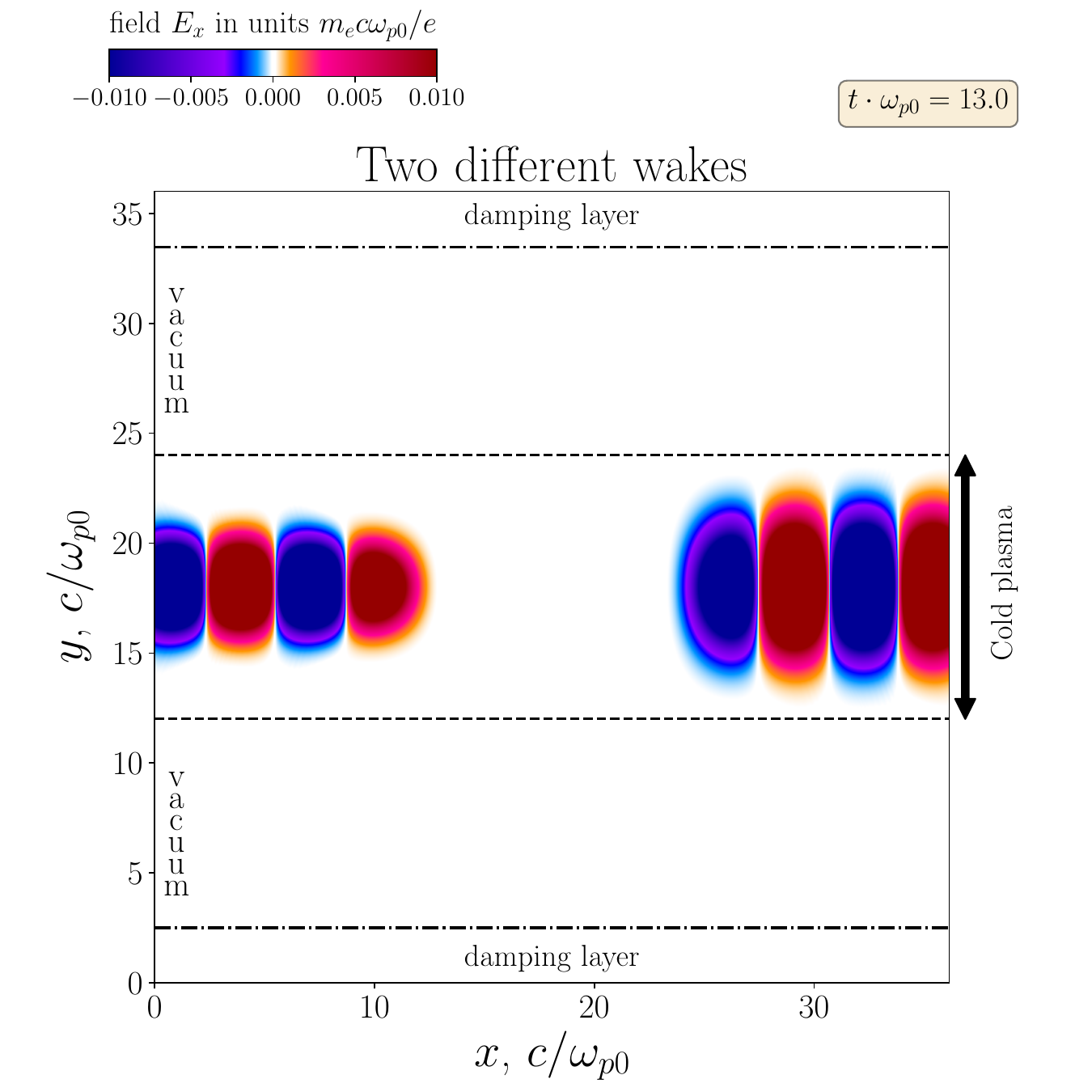}	
			\includegraphics[width=1\linewidth]{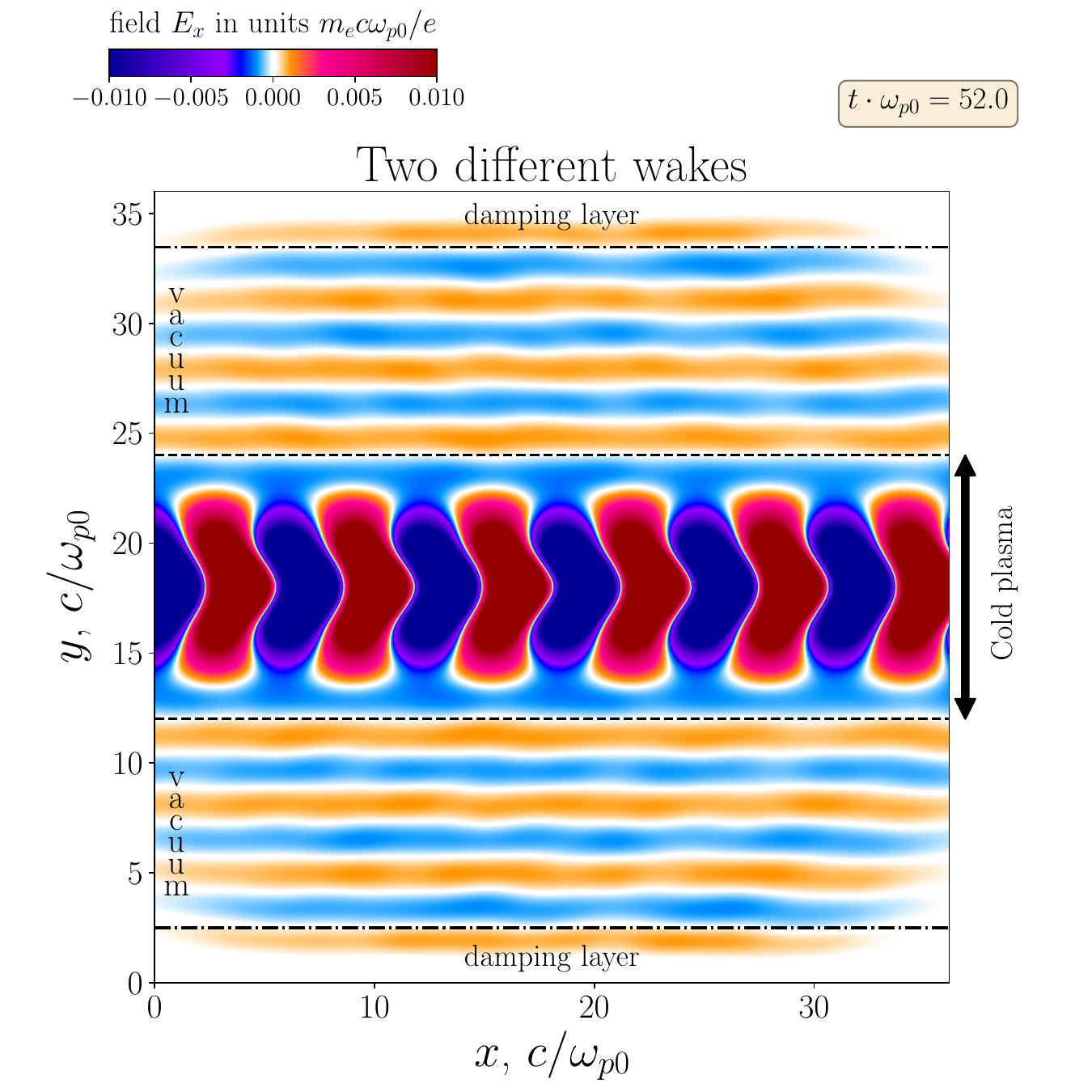}
			
			\caption{Counter plasma waves with the different transverse shapes.}\label{fig:diffWakes}
		\end{minipage}
	\end{center}
\end{figure*}

The mechanism of radiation due to interaction of counter plasma waves is described in detail in \cite{Timofeev2017b}, so in this section we will limit ourselves to a brief description of its basic essence. Let us consider counter-propagation of two plasma waves in a cold and unmagnetized plasma. Their amplitudes are $E_1(\mathbf{r}_\perp)$ and $E_2(\mathbf{r}_\perp)$, where $\mathbf{r}_\perp$ is the coordinate transverse to the direction of wave propagation. The symbol $\parallel$ will denote the components along the line of wave propagation. The scattering of one wave on the density perturbation created by another wave generates a perturbation of plasma electron current:

	\begin{eqnarray}\label{eq:current}
		\mathcal{J}=\frac{1}{4}\left[(k_\parallel^{(1)}-k_\parallel^{(2)})\left(2E_1E_2+\frac{1}{k_\parallel^{(1)}k_\parallel^{(2)}}E_1^{\prime}E_2^{\prime}\right)+\right.\\
		\left.+\left(\frac{1}{k_\parallel^{(2)}}E_1E_2^{\prime\prime}-\frac{1}{k_\parallel^{(1)}}E_1^{\prime\prime}E_2\right)\right],\nonumber
	\end{eqnarray} 

here $k_\parallel^{(1)}$ and $k_\parallel^{(2)}$ are the longitudinal components of wave vectors of the plasma waves and $^{\prime}\equiv{\partial}/{\partial \mathbf{r}_\perp}$ is the derivative in transverse direction. The frequency of this perturbation will be equal to the sum of the frequencies of the original waves $\omega_r=\omega_1+\omega_2$ and the wave number will be equal to the sum of the colliding wave numbers $k_\parallel^{(r)}=k_\parallel^{(1)}+(-k_\parallel^{(2)})$. Such a perturbation can have a superluminal phase velocity $v_{ph}={\omega_r}/{k_\parallel^{(r)}}$ and is therefore able to excite electromagnetic oscillations at the frequency $\omega_r$ in a resonant manner. In the particular case of colliding waves with identical wave vectors, the longitudinal wave number of the resulting perturbation is zero and its phase velocity becomes infinite.

The first term in \ref{eq:current} depends substantially on the wave numbers of the colliding waves. The wave number $k$ is determined by the velocity of the drivers. So, first term is smaller the closer to each other velocities of the drivers. Thus, for example, it is exactly zero for waves excited by short laser pulses (traveling at the speed of light). The second term can contribute to the radiative current if the transverse profiles of the colliding oscillations are different ($E^{\prime\prime}_1({\bf r}_{\bot})\neq E^{\prime\prime}_2({\bf r}_{\bot})$). According to the theory \cite{Timofeev2017b} the most efficient emission is expected when the cross size of the colliding plasma waves is of the order of $c/\omega_p$.

Let us demonstrate this effect using synthetic calculations of propagation in cold unmagnetized plasma of counter plasma waves generated by the ponderomotive force of short laser pulses. For simplicity, we turn off the effect of laser diffraction and ion dynamic. The figure \ref{fig:sameWakes} shows the case of plasma waves with the same transverse structure. One can see the absence of radiation before the collision of the pulses, which is an obvious consequence of the potential nature of the plasma waves. After the overlap of the waves with the same shapes a slight radiation is observed, which quickly ceases. In the case of plasma waves with different transverse structure (figure \ref{fig:diffWakes}), a significant emission at the second harmonic of the plasma frequency in the strictly transverse direction is seen. More details on these calculations can be found in the video in the Supplementary materials.

For radiation at the second harmonic of the plasma frequency, the plasma is actually transparent. Thus we can use for the EM waves the dispersion relation: $\omega_r^2= k^2c^2$, where  $k^2=k_\parallel^{(r)2}+k_\perp^{(r)2}$ and $\omega_r$ is the frequency of radiation. In the case of different wave numbers $k_\parallel^{(1)}$ and $k_\parallel^{(2)}$ of colliding waves, the radiation will have a longitudinal wave number $k_\parallel^{(r)}=k_\parallel^{(1)}-k_\parallel^{(2)}$ and therefore will exit at some angle. The longitudinal wave number and frequency of the generated radiation is determined by the parameters of the colliding waves, so it is not difficult to calculate the resulting transverse wave number and angle of emission:
\begin{equation}\label{eq:angle}
	\alpha=\arctan\left[\dfrac{k_\perp^{(r)}}{k_\parallel^{(r)}}\right],
\end{equation} where $k_\perp^{(r)} c=\sqrt{\omega_r^2-k_\parallel^{(r)2} c^2}$.

\section{Beams and plasma parameters}\label{sec:param}

First of all, it is necessary to choose such parameters of the beam-plasma system at which oblique oscillations with a transverse wavelength of the order of units $c/\omega_p$ are the most unstable. Dominance of such oscillations in the unstable spectrum is primarily a result of relativistic speed of the beam particles. In this case due to relativistic mass anisotropy the oscillations transverse to the beam propagation direction appear to be easier to excite. The main obstacle to the excitation of oblique oscillations is the presence of an external magnetic field directed along the beam propagation axis. What kind of oscillations will be excited in a given beam-plasma system can be estimated from the linear theory. In the case of relativistic beams in a weak or zero-point magnetic field without taking kinetic effects into account (cold plasma), a typical instability growth rate is an almost continuously increasing function of the transverse wave number $k_\perp$. Taking into account the temperature spreads of the beam particles has a significant effect on this dependence. The oblique instabilities \cite{Timofeev2013b} are primarily suppressed with increasing temperature.

In \cite{Annenkov2020} it has been shown that even when simulating a realistic beam injection into plasma, at the linear stage of instability plasma oscillations are excited with a spectrum in excellent agreement with predictions of linear theory, taking into account kinetic and relativistic effects, as well as the presence of an external magnetic field. In this paper, we consider beams with velocities $v_b=0.9c$ and $v_b=0.7c$, where $c$ is the speed of light.  Using the DispLib \cite{Annenkov2021b} library we have chosen regimes for them at which the transverse wave number of the most unstable modes $k_\perp^{max}=2$~$\omega_p/c$. The first mode ($v_b=0.9c$) we will further call relativistic. The second mode with $v_b=0.7c$ we will further call subrelativistic. Both beams have a Maxwellian energy spread $$f_b(\textbf{p})\propto\exp\left(-\dfrac{(\textbf{p}-\textbf{p}_0)^2}{2\Delta p_b^2}\right),$$ and temperature is defined as $T_b=\Delta p_b^2/(2m_e)$.

Further we will operate with dimensionless values. Plasma and beam particle densities will be calculated in units of $n_0$; all frequencies in $\omega_{p0}=\sqrt{{4\pi n_0 e^2}/{m_e}}$; wave numbers in $\omega_{p0}/c$; lengths in $c/\omega_{p0}$; time in $\omega_{p0}^{-1}$; electromagnetic fields measured in units of $m_ec\omega_{p0}/c$; particle speeds in the speed of light $c$.

\begin{table}
	\begin{center}
	\caption{Parameters of the beam-plasma system.}\label{tab:params}
	
	\begin{tabular}{c|c|c|c|c|c|c}
		
		$v_b/c$	& $n_b/n_0$ & $T_b$, keV &   	$\Gamma/\omega_{p0}$ & $\omega_b/\omega_{p0}$ & $k_\perp^{max}$, $\omega_{p0}/c$		&  $k_\parallel^{max}$, $\omega_{p0}/c$		  \\[6pt]
	\hline	
		0.9 & 0.01 &  1 &  0.09 & 0.94 & 2& 1.14 \\
		
		0.7 & 0.03 &  0.75 &  0.154 & 0.87 & 2& 1.46 \\
		
	\end{tabular}
\end{center}
\end{table}
\begin{figure*}
	\centering
	\includegraphics[width=0.85\linewidth]{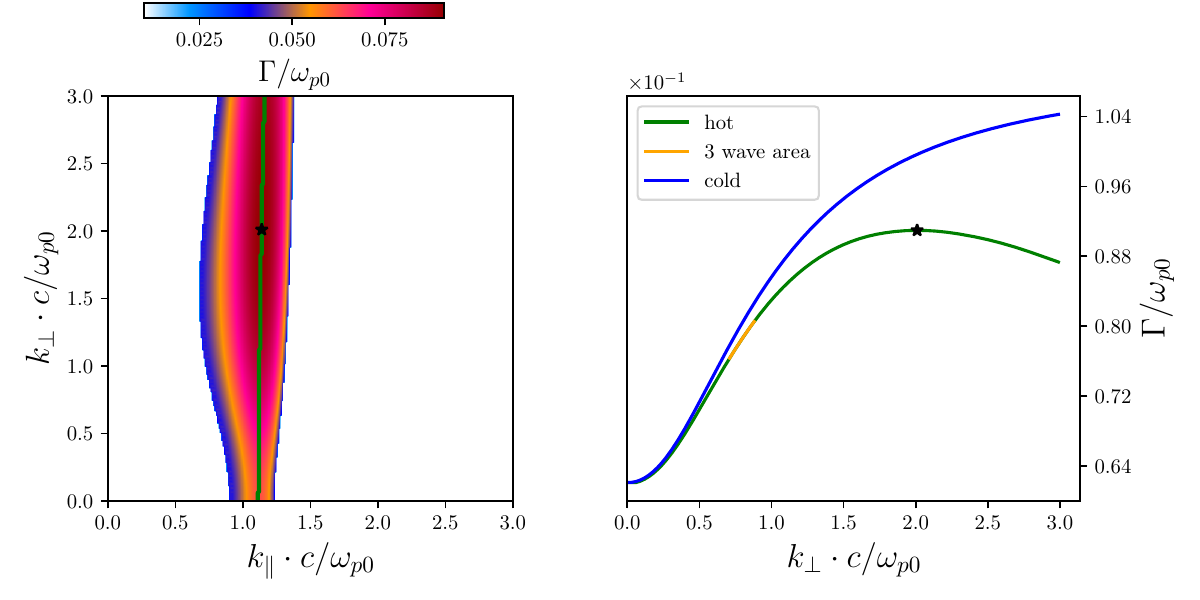}
	\caption{Left: The growth rate map for the beam-plasma instability $\Gamma\left(k_\parallel,k_\perp\right)$  for the beam with $v_b=0.9c$.   The green line $k_\perp=k_\perp(k_\parallel)$ marks the local maximal growth rate. Right: $\Gamma(k_\perp)$ is on the line of the maximum (orange line indicates the  region of the three-wave interaction). Blue line corresponds to the case of cold beam and plasma.}\label{fig:inc09}
	\includegraphics[width=0.85\linewidth]{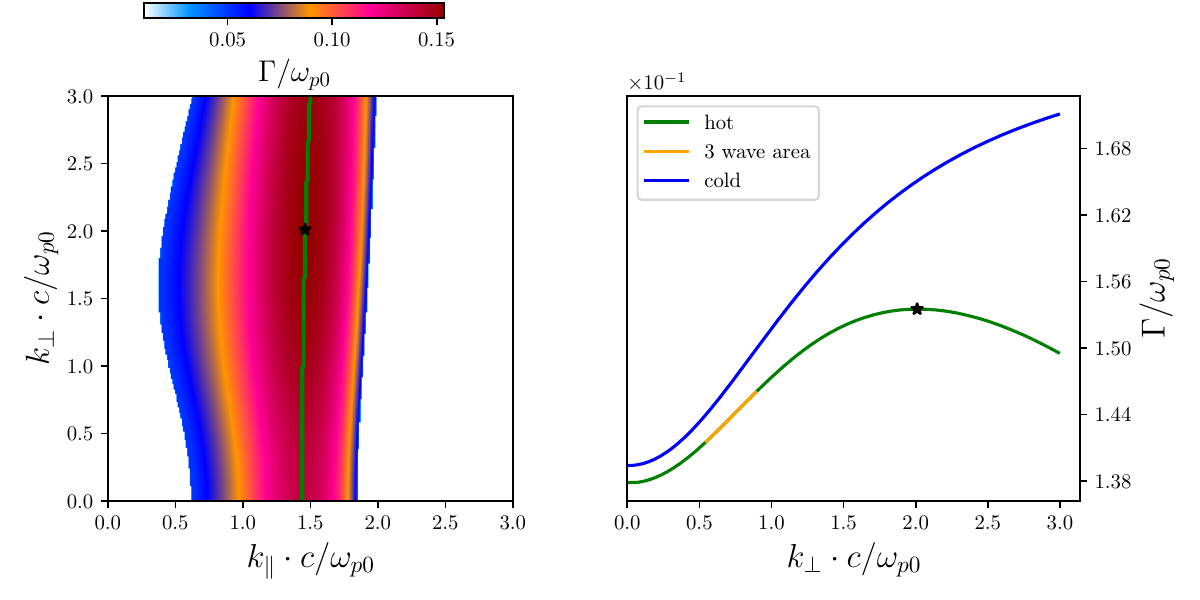}
	\caption{Same as on the figure \ref{fig:inc09}, but for the beam with $v_b=0.7c$.}
	\label{fig:inc07}
\end{figure*}
Table \ref{tab:params} shows the parameters of the selected regimes, as well as the characteristics of the most unstable plasma beam modes: the excitation frequency $\omega_b$, the wave numbers $\mathbf{k}=(k_\parallel,k_\perp,0)$ and the growth rate value $\Gamma$. First, we chose parameters of relativistic beam. Then we adjusted density and temperature of the subrelativistic beam so that  the transverse number of the most unstable mode also was $k_\perp=2$. Since decreasing the beam velocity reduces the growth rate  of oblique instabilities, we had to raise the relative density of the beam and decrease its temperature. The resulting growth rate is higher than in the case of relativistic beam. This means that the relaxation length of such a beam will be smaller. One can also expect a larger amplitude of excited plasma oscillations in this case. In both modes, the background plasma temperature was $T_p=50$ eV. This temperature has no significant effect on the growth rate of the instability of such beams but is necessary for the stability of the numerical schemes in subsequent simulations. Also the whole system is immersed in an external magnetic field $\textbf{B}=(B_0,0,0)$ such that $\Omega_e=0.1\omega_p$, where $\Omega_e=|e|B_0/m_e c$ is the cyclotron frequency of electrons.

 Figures \ref{fig:inc09} and \ref{fig:inc07} show the results of calculation of the growth rate  for relativistic and subrelativistic beams, respectively.  For comparison, the line of maximum growth rate for the model with cold plasma is also shown. Orange color  marks  the region of spectrum, where oscillations in case of counter beams can participate in three-wave process of two Langmuir waves coupling into electromagnetic one with frequency equal to the doubled plasma frequency. In \cite{Annenkov2020} it has been shown that if the most unstable modes lie in this region, then, already at the linear stage of the beam-plasma instability, a highly efficient generation of electromagnetic radiation is possible. However, even a relatively small change of the system parameters, leading to violation of such localization of the maximum growth rate leads to a significant decrease of the emission level. Therefore, for the parameters chosen in this paper, we can be sure that not a three-wave process will determine the emission process of EM waves in the system.

\section{Simulation setup}\label{sec:setup}

For numerical simulations, we use our own 2D3V Cartesian parallel PIC code implemented for Nvidia GPGPU \cite{Lindholm2008}. It is based on standard computational schemes: the  \cite{Yee1966} solver of Maxwell equations for EM fields, the \cite{Boris1970} scheme for solving the equation of motion for collisonless macro-particles with a parabolic form factor, and the charge-conserving \cite{Esirkepov2001} scheme  for calculations of currents. In all calculations, the spatial step of the computational grid was \begin{figure}
	\centering
	\includegraphics[width=\linewidth]{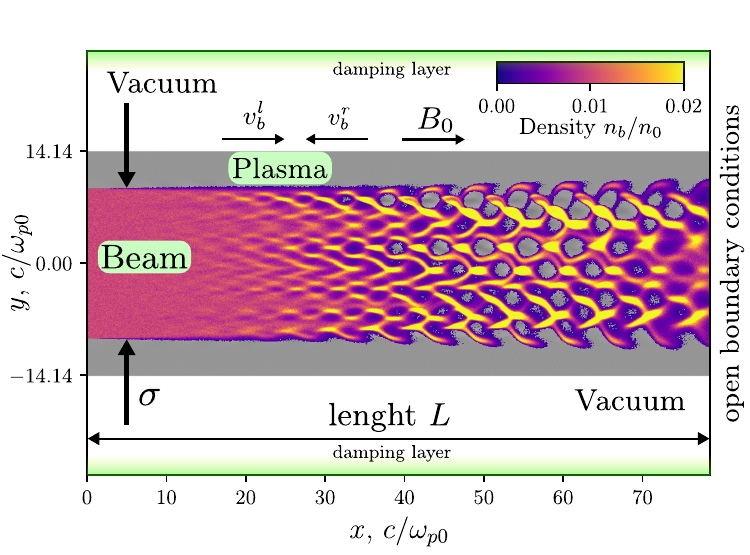}
	\caption{Schematic of the computational region.}
	\label{fig:sch}
\end{figure}
$\Delta x=\Delta y=\pi/80\approx0.04$, and the time step was $\Delta t=0.5\Delta x$.

Figure \ref{fig:sch} shows a schematic of the computational region. At the center of the computational domain a plasma column with a width of $720$ cells ($\approx28.3$ $c/\omega_{p0}$) is located, the plasma length $L$ will be varied for different regimes. Open boundary conditions are used at the ends of the column. They allow us to simulate the continuous injection of electron beams by self-consistent maintaining the compensating plasma current arising from the electron beam propagation. Figure \ref{fig:sch} shows the density of one relativistic beam at the moment of time shortly after it reaches the opposite boundary. The transverse structure of the density perturbation arising due to the oblique instability is clearly seen. The colliding beams will have the same width $\sigma$, which we will vary in calculations from $3$ to $9$ transverse wavelengths $\lambda_\perp={2\pi}/{k_\perp^{max}}=\pi$.  The injected beam current grows smoothly from zero to a given value in time $50$~$\omega_{p0}^{-1}$ to avoid creating a seed to excite purely longitudinal oscillations by a sharp beam front \cite{Volchok2019b}.  The plasma is surrounded by a vacuum and EM absorbing layers are located at the boundary. A description of the implementation of open boundary conditions and absorbing layers is available in \cite{Annenkov2018}. We will not consider the dynamics of plasma ions in this study, as it will interfere with a pure consideration of the process of interest.

\section{Simulation results}\label{sec:result}

Let us first consider the emission process in a system of counter-streaming relativistic beams.
\subsection{Relativistic regime}\label{subsec:rel}

Before proceeding to counter-beam simulations, two important issues have to be resolved. Firstly, it is necessary to determine a sufficient number of macro-particles in the cell for the PIC simulation. Secondly, to find out at what distance from the beam injection site there is a region of beam relaxation where the most intensive plasma oscillations will be excited. This is necessary to choose the longitudinal size $L$ of the computational region. For the most efficient emission the beams should relax in one place, so the length $L$ should be approximately equal to twice the relaxation length of a single beam. Both questions can be answered by simulating single beam injection.

\subsubsection{Single beam}\label{subsec:single09}

Since in considering the relaxation process we are not interested in emission processes, it is possible to simplify the computational domain by removing vacuum gaps, imposing periodic boundary conditions in transverse direction and reducing the thickness of the beam-plasma system to a few $\lambda_\perp$. This makes it possible to reduce the computational resource requirements and simplifies the study of the convergence of calculations on the number of particles per cell (ppc). In these calculations we will be interested in two quantities. The first is the beam instability growth rate, which is calculated from the simulation results as follows:
\begin{align*}
	\Gamma^{PIC}(t)=\dfrac{H'(t)}{2H(t)},
\end{align*}
where $H(t)$ is the energy of the electric field in the whole area.

The second quantity is the positioning of the beam relaxation region (the region of most intense excitation of plasma oscillations). Since we have to take into account that the beam instability is oblique, we will trace the location and amplitude of the quantity $E_0$ \cite{Annenkov2020}, which is calculated as the amplitude of plasma oscillations averaged over the length of plasma oscillations $2l=2\pi c/\omega_p$ in the longitudinal direction and integrated over the plasma thickness $L_y$:
\begin{gather}
	E_0(x,t)=\int\limits_{0}^{L_y}dy\left[\dfrac{\omega_p}{c\pi}\int\limits_{-l}^{l}dx'E_x^2(x+x',y,t)\right]^{1/2}.
\end{gather}
After calculation of $E_0(x,t)$ at each diagnostic time step, the value of the amplitude maximum as well as its longitudinal position are determined.

\begin{figure*}
	\centering
	\includegraphics[width=1\linewidth]{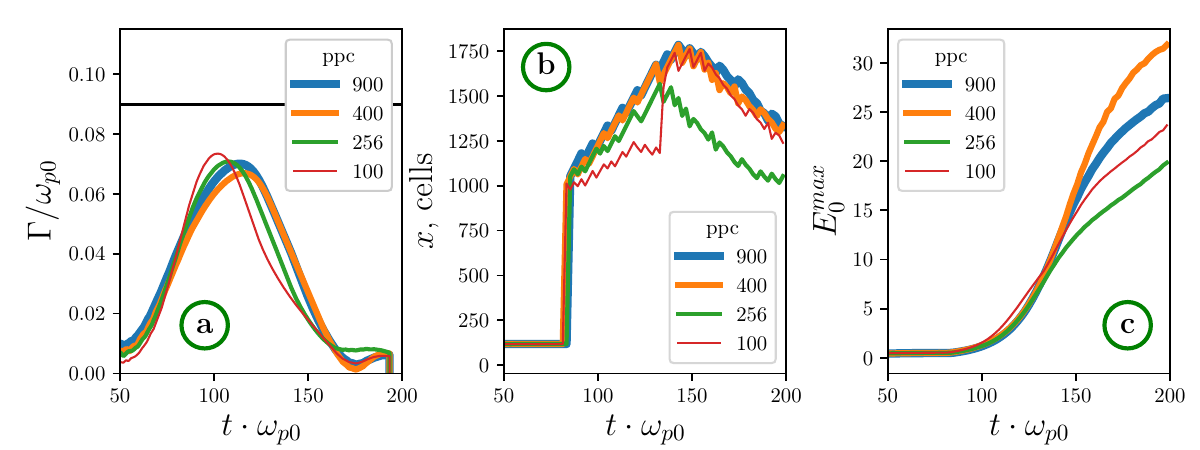}
	\caption{Simulation of a single ''relativistic'' beam injection into plasma. The thickness of the system is $3\pi$. In transverse direction periodic boundary conditions are realized. a) Dependence of the instability growth rate $\Gamma$ on time. (b) Dependence of the longitudinal coordinate of the maximum value of $E_0(x,t)$. (c) Dependence of the magnitude of the maximum $E_0(x,t)$.}
\label{fig:inc_pic09}
\end{figure*}
\begin{figure*}
	\centering
	\includegraphics[width=0.48\linewidth]{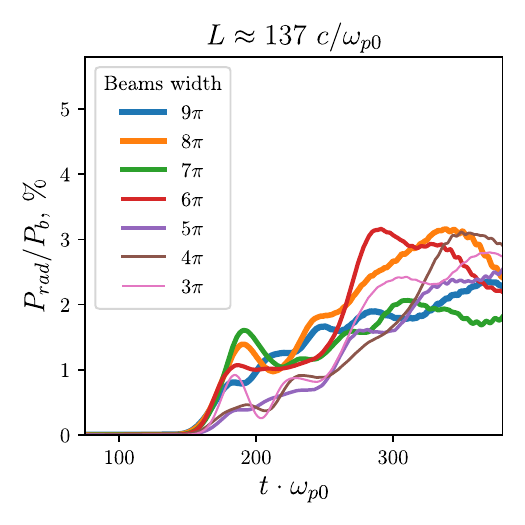}
	\hfill
	\includegraphics[width=0.48\linewidth]{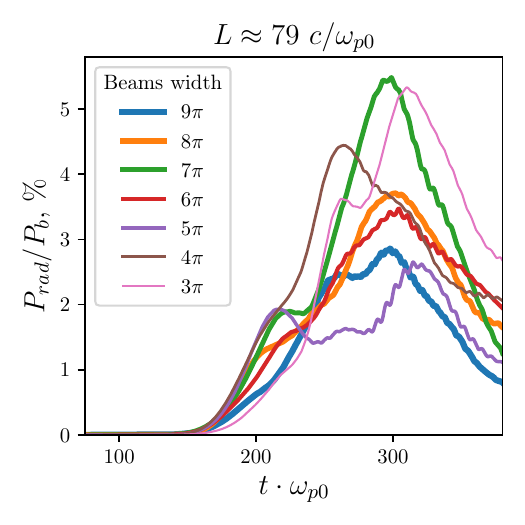}
	\caption{Efficiency of beam-to-radiation power conversion as a function of time for relativistic counter-streaming beams.  Left: \textit{long} system, right: \textit{short} system. The plots are shown for all considered beam thicknesses.}
	\label{fig:eff09}
\end{figure*}

Single beam relaxation in plasma was simulated at system transverse sizes $1\lambda_\perp$, $2\lambda_\perp$ and $3\lambda_\perp$ (results were identical) for number of particles of each kind in a cell $100$, $256$, $400$ and $900$. Figure \ref{fig:inc_pic09} shows the results. It can be seen that as the number of particles increases, there is no convergence to a single value. The observed differences seem to be caused by the dependence of the beam relaxation scenario on a particular implementation of the macro-particle distribution function which is natural for PIC calculations of the beam-plasma interaction \cite{Annenkov2020}. The absolute value of the instability growth rate in the beam injection simulations turned out to be somewhat smaller than theoretical predictions for an infinite beam-plasma system. Since $100$ ppc allows us, with the available computational power, to investigate a larger range of thicknesses of the beam-plasma system, further calculations will be carried out with this number of ppc.  Based on the dependence of $E_0$ on the coordinate, we have chosen two lengths: $L=2\cdot1750\Delta x\approx137.5$~$c/\omega_{p0}$ (\textit{long}) and $L=2\cdot1000\Delta x\approx78.5$~$c/\omega_{p0}$ (\textit{short}).

Let us note that this is an approximate method of determining the optimal system size for radiation generation. Obviously, in a system with two beams, their relaxation follows a slightly different scenario from the one-beam case. The higher the relative density $n_b$ of the beams and the further the two-beam instability process is from the linear regime, the stronger this difference will be. Therefore if the goal is to maximise the level of EM emission, multiple simulations of different distances between the beam injection sites are required.

\begin{figure*}
	\centering
	\includegraphics[width=1\linewidth]{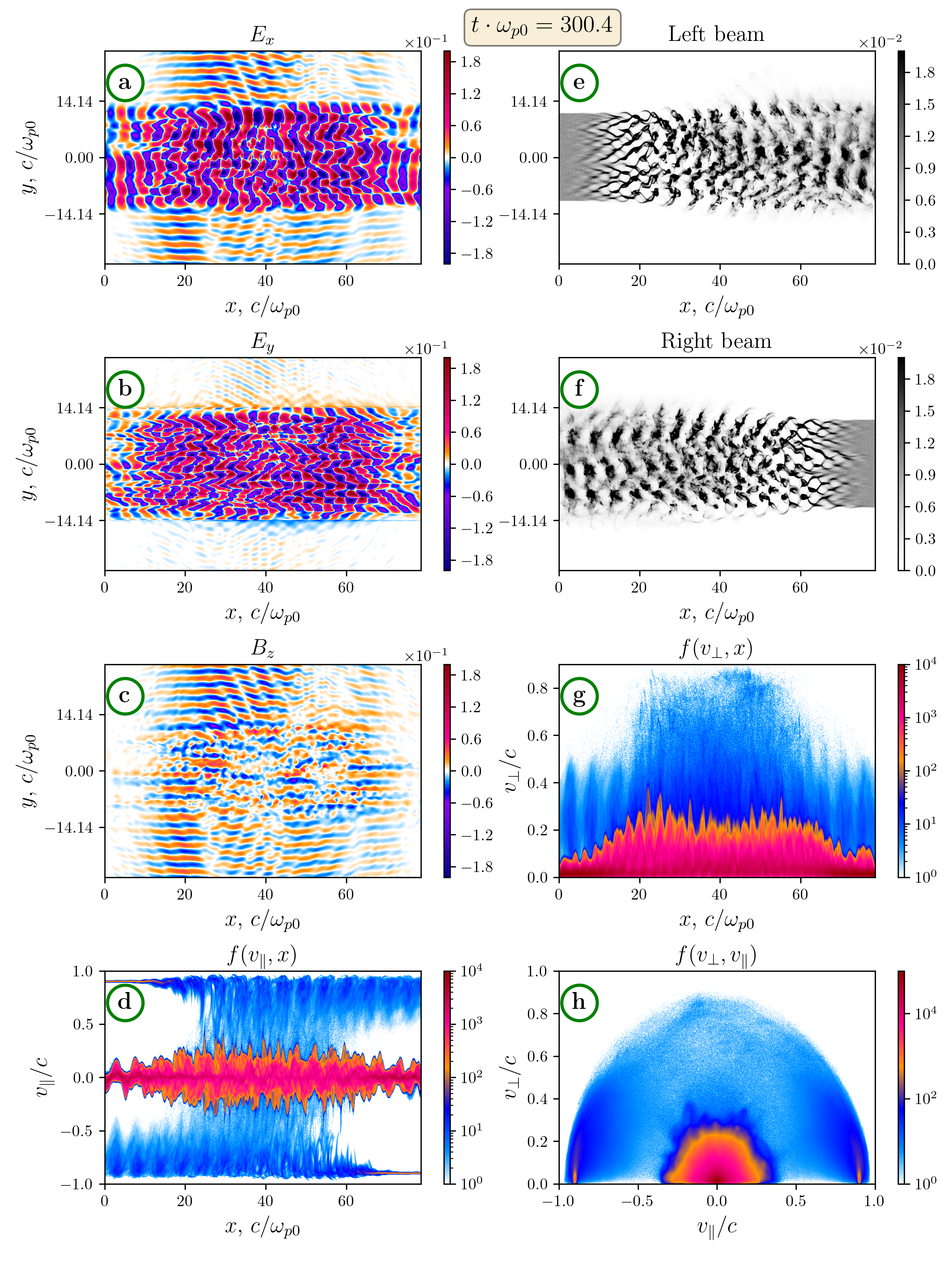}
	\caption{Simulation results for relativistic counter-streaming beams. a)-c) -- components of electromagnetic fields. e) and f) -- beams density. d), g) and h) -- plasma and beams electron velocity distribution functions. Beams thickness  is $\sigma=7\pi$, time moment: $t\cdot\omega_{p0}\approx300$.}
	\label{fig:pic09}
\end{figure*}

\begin{figure*}
	\centering
	\includegraphics[width=1\linewidth]{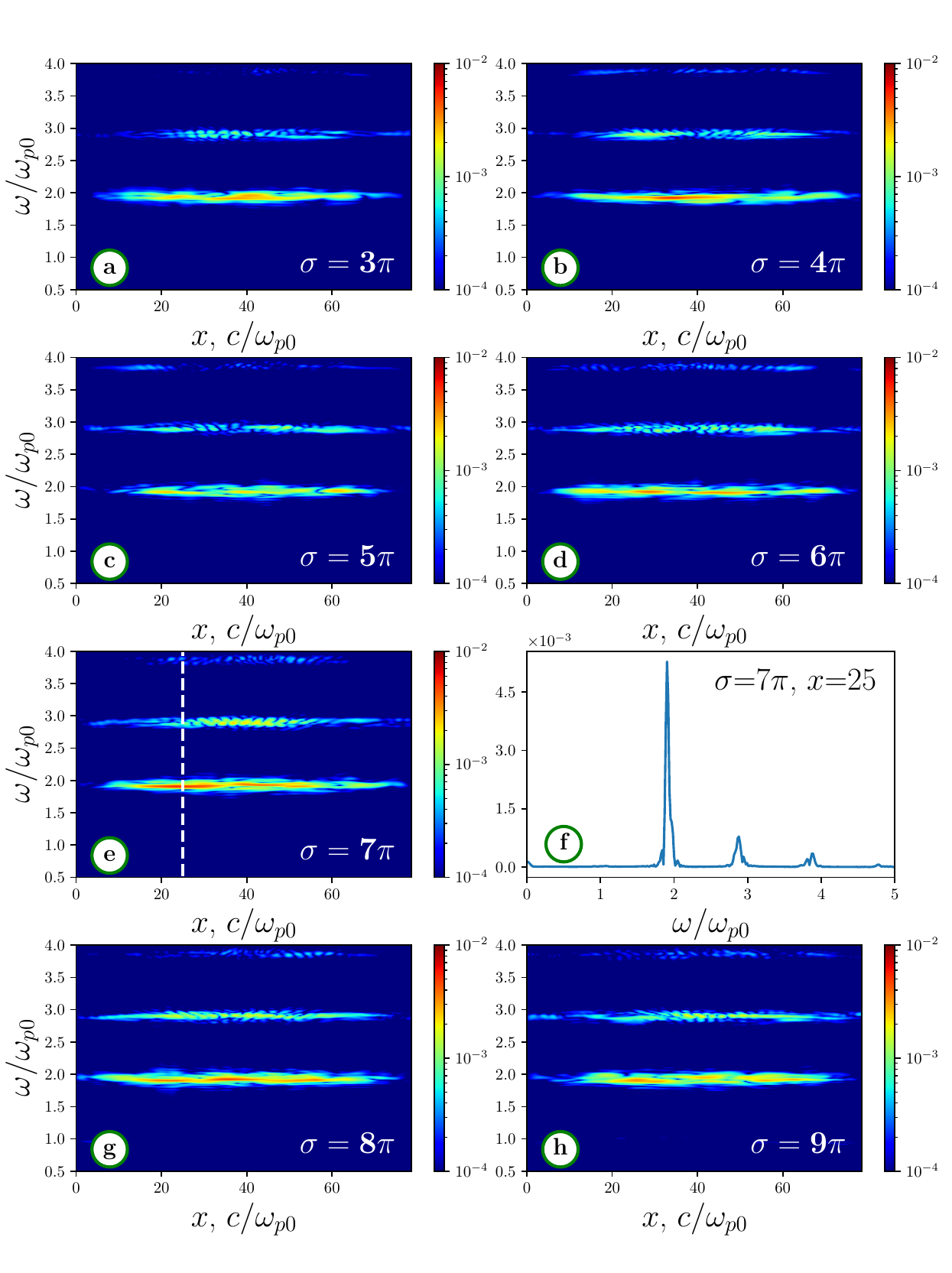}
	\caption{Fourier spectrum of emission as a function of coordinate $x$ for all $\sigma$ in case of counter-streaming relativistic beams. f) -- spectrum in $x=25c/\omega_{p0}$ for $\sigma=7\pi$.}
	\label{fig:fft09}
\end{figure*}

\begin{figure*}[t]
	\centering
	\includegraphics[width=1\linewidth]{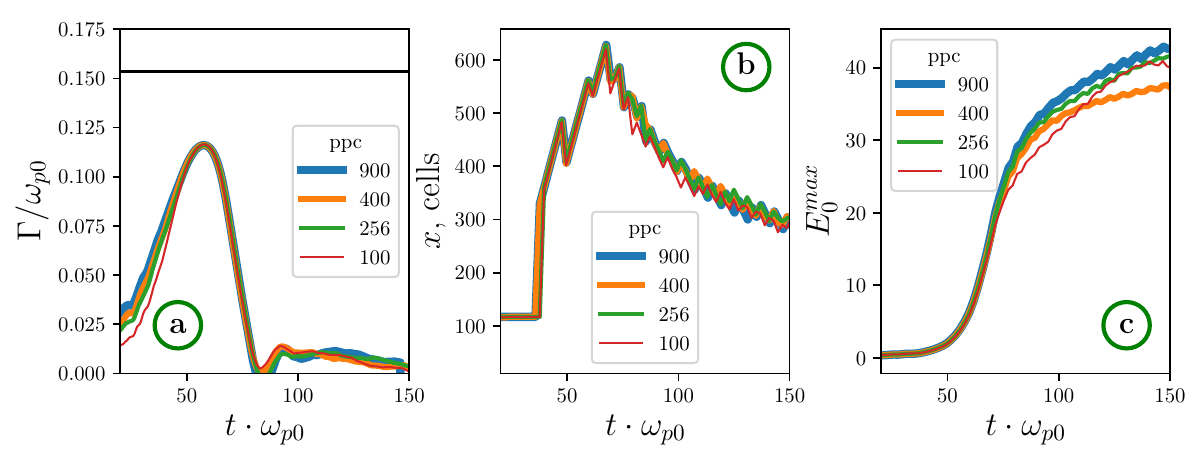}
	\caption{Same as on Figure \ref{fig:inc_pic09}, but for sub-relativistic beam.}
	\label{fig:inc_pic07}
\end{figure*}

\subsubsection{Counterstreaming beams}\label{subsec:counter09}

Figure \ref{fig:pic09} shows simulation results for the injection of counter-streaming relativistic beams with transverse size $\sigma=7\pi$ into a \textit{short} plasma. Intense transverse emission can be seen from the region where the relaxation regions of each beams intersect. One can estimate their location by looking at the phase portrait $f(v_\parallel,x)$ (figure \ref{fig:pic09}d). Figure \ref{fig:pic09}b also shows the presence in the system of a small emission at a higher frequency. The radiation spectrum as a function of the longitudinal coordinate for each beam thickness is analyzed in detail in figure \ref{fig:fft09}. It can be seen that there is no significant difference for each case. Apart from the main radiation near the second harmonic of the plasma frequency \begin{equation}\label{coal}
	(\omega_b,k_\parallel)+(\omega_b,-k_\parallel)\rightarrow (2\omega_b,0),
\end{equation} there is also a small amount of radiation near the third and fourth harmonics, arising from the processes  \begin{align}
&(\omega_b,\mp k_\parallel)+(2\omega_b,\pm 2k_\parallel)\rightarrow (3\omega_b,\pm k_\parallel), \label{p1} \\
&(2\omega_b,\pm 2k_\parallel)+(2\omega_b,0)\rightarrow (4\omega_b,\pm 2k_\parallel). \label{p5}
\end{align} These processes are essentially non-linear and may be relevant for beams with high relative density that excite plasma oscillations of large amplitude, as for example in \cite{Annenkov2021}.

Figure \ref{fig:eff09} shows the beam-to-radiation power conversion efficiency. It can be seen that it reaches a value of a few percent and has no obvious dependence on thickness of the injected beams. However, in the case of \textit{short} plasma the radiation process is more efficient. One possible reason for this is that the emission process begins at the initial stage of instability, which is closer to linear. On this stage, there is a more distinct transverse structure of the excited waves. At large plasma lengths the emission is generated from regions distant from the place of development of linear instability. Reaching them, the beams have a prehistory of interaction with the developed fields in the plasma and the oscillations excited by them have a transverse structure which is less suitable for the generation of radiation.

\subsection{Sub-relativistic regime}\label{subsec:subrel}

Let us show that emission generation by the discussed mechanism for subrelativistic beams is realized similarly to relativistic beams. As in the previous section, we first will investigate the convergence on the number of particles per cell and also we will estimate the relaxation length of a single beam by simulating a beam-plasma system with periodic boundary conditions in transverse direction and small thickness.
\subsubsection{Single beam}\label{subsec:single07}

Figure \ref{fig:inc_pic07} shows the time dependence of the instability growth rate $\Gamma(t)$ obtained in numerical simulations of single beam injection into plasma with different ppc, as well as the localization and magnitude of the maximum $E_0(x,t)$. As in the case of the relativistic beam, no fundamental difference in the results has been observed for beam-plasma systems of different thicknesses ($1\lambda_\perp$, $2\lambda_\perp$ and $3\lambda_\perp$). Similarly to the previous regime, the maximum value of the growth rate is slightly lower than predicted by the linear theory. But for this case, there is in fact a complete coincidence of the obtained dependencies for different ppc. From this it is possible to conclude that for the chosen beam parameters the development of instability is actually in hydrodynamic regime and less sensitive to details of reconstruction of distribution function by macro-particles.

\begin{figure*}[t]
	\centering
	\includegraphics[width=\linewidth]{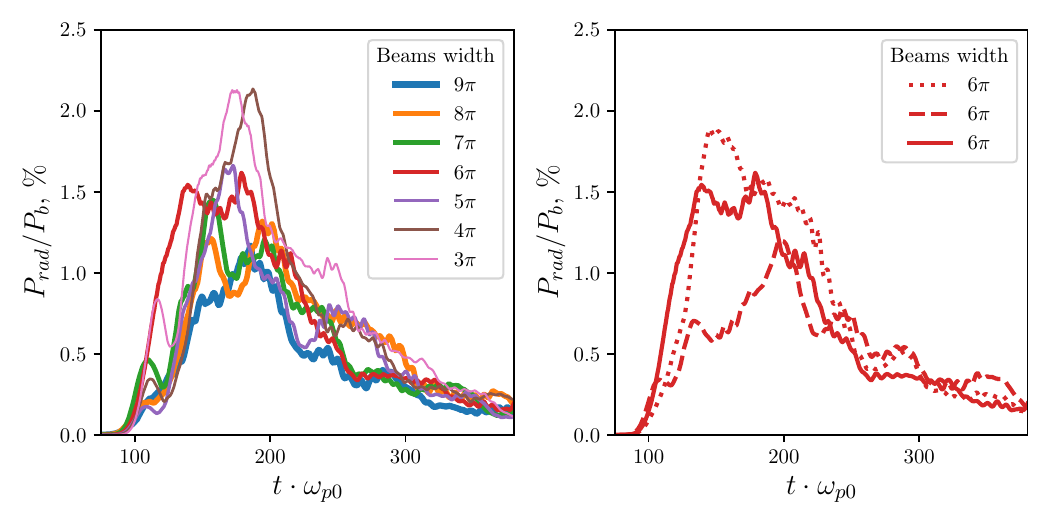}
	\caption{Efficiency of beam-to-radiation power conversion as a function of time for sub-relativistic counter-streaming beams.  Left: all beam thicknesses except $\sigma=6\pi$, right: three calculations with $\sigma=6\pi$.}
	\label{fig:eff07}
\end{figure*}

For the considered sub-relativistic beam, the relaxation length is shorter than for the relativistic beam. The development of instability occurs earlier because of the larger growth rate. On the basis of obtained results, the beam-plasma system length for calculations with counter beams has been chosen equal to $L=2\cdot450\Delta x\approx35.3$~$c/\omega_{p0}$.

\subsubsection{Counterstreaming beams}\label{subsec:counter07}

\begin{figure*}
	\centering
	\includegraphics[width=1\linewidth]{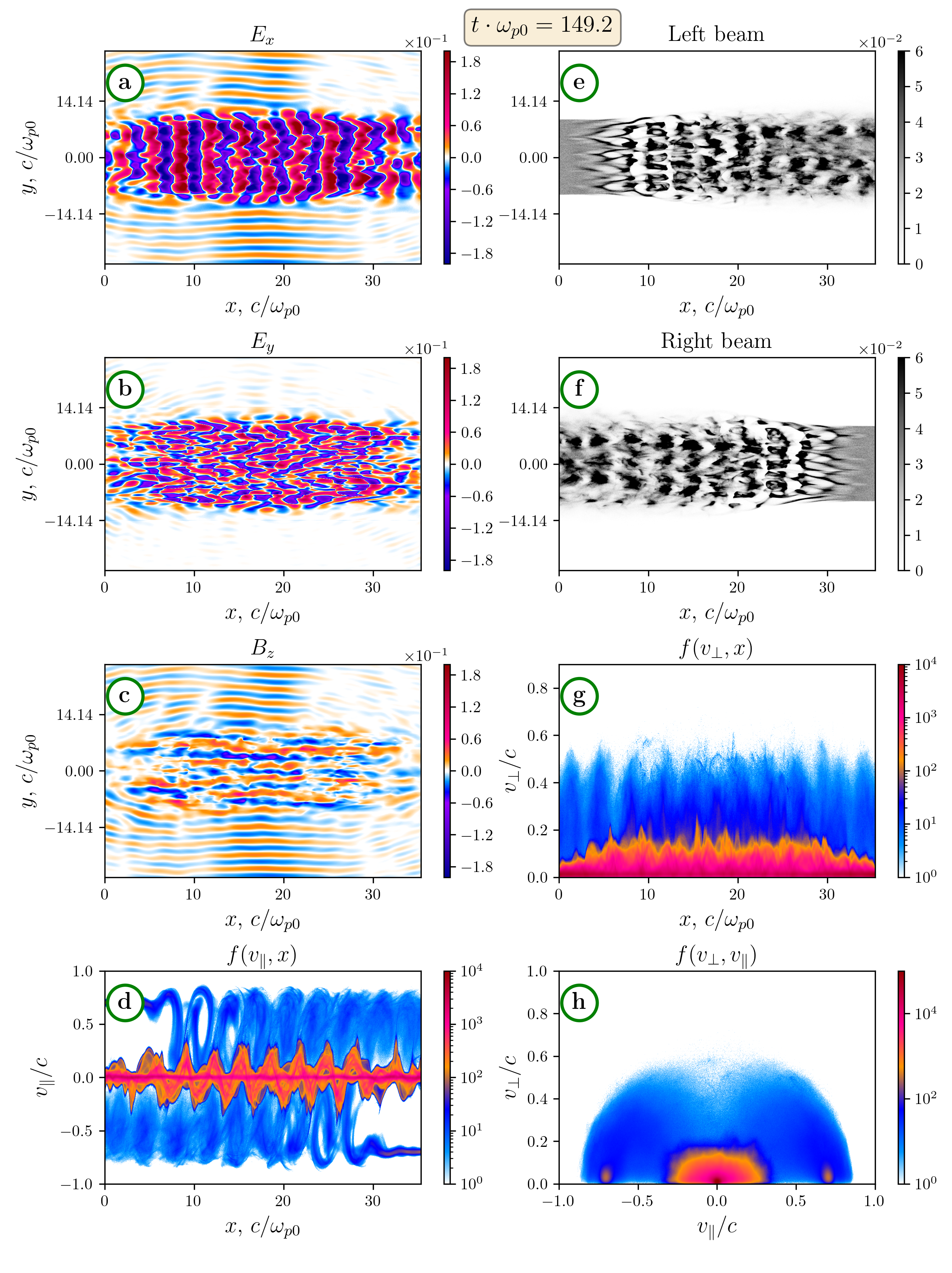}
	\caption{Same as on fig. \ref{fig:pic09}. Sub-relativistic beams with thickness $\sigma=6\pi$. Time moment $t\cdot\omega_{p0}\approx150$}
	\label{fig:pic07}
\end{figure*}

Similarly as for the relativistic regime we investigate the dependence of the radiation parameters on the transverse size of the injected beams. Figure \ref{fig:pic07} shows simulation results for beam widths $\sigma=6\lambda_\perp$. 

\begin{figure}
	\centering
	\includegraphics[width=\linewidth]{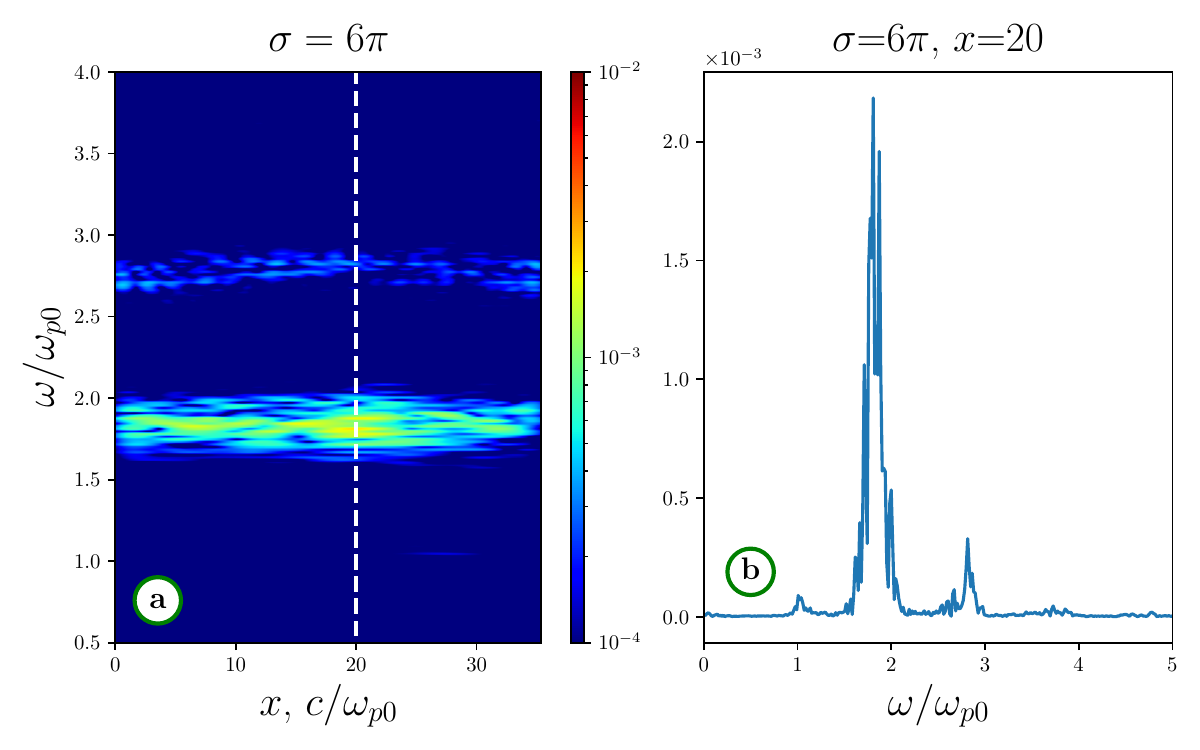}
	\caption{Fourier spectrum of emission for sub-relativistic counter-streaming beams with  $\sigma=6\pi$.}
	\label{fig:fft07}
\end{figure}

Figure \ref{fig:eff07} shows the histories of the radiation efficiency for different beam thicknesses. The result presented in the left figure looks as if some explicit dependence on the beam thicknesses takes place. To check this, we have carried out three simulations with $\sigma=6\lambda_\perp$ beams (right figure) and different actual realizations of the macro-particle distribution. It can be seen from the results that the observed differences in the radiation efficiency are related to these realizations and there is no clear dependence on the thickness of the beams. The maximum efficiency achieved is somewhat lower than in the relativistic mode. There can be different reasons for this, for example, smaller length of emitting region or more ''flat form'' of excited plasma oscillations, caused by development of more intensive instability.

Figure \ref{fig:fft07} shows the emission spectrum over the whole computation time for beams with $\sigma=6\lambda_\perp$. In this case, a slightly broader radiation is observed which is also caused by the development of a more intense instability changing the parameters of the beam-plasma system and decreasing the emission duration.

\subsection{Beams with different velocities}\label{subsec:diffbeams}

\begin{figure}
	
	\begin{minipage}[h]{\linewidth}
		\includegraphics[width=1\linewidth]{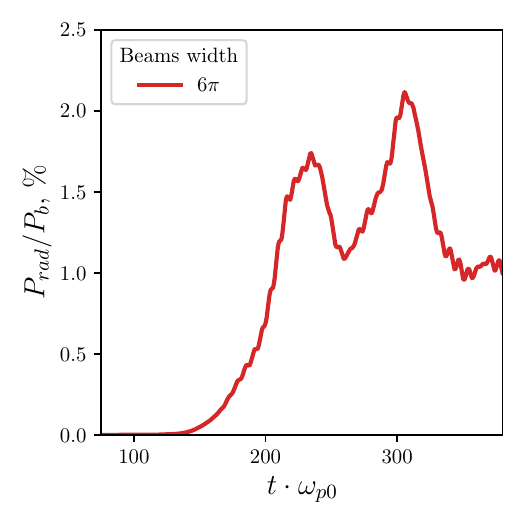}
		\caption{Efficiency of beam-to-radiation power conversion as a function of time for counter-streaming relativistic and sub-relativistic beams. }
		\label{fig:eff0709}
	\end{minipage}
\end{figure}

\begin{figure}
	\begin{minipage}[h]{\linewidth}
		\centering
		\includegraphics[width=1\linewidth]{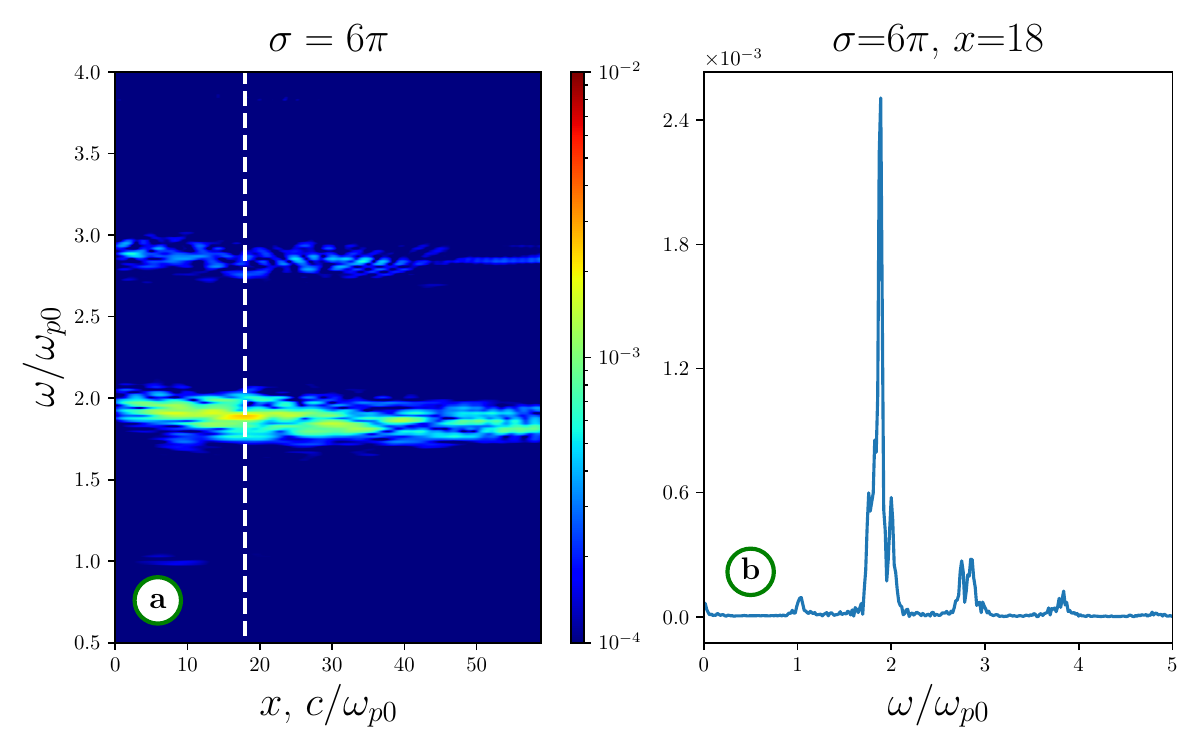}
		\caption{Emission spectrum for counter-streaming relativistic and sub-relativistic beams.}
		\label{fig:fft0709}
	\end{minipage}
\end{figure}

\begin{figure*}
	\centering
	\includegraphics[width=1\linewidth]{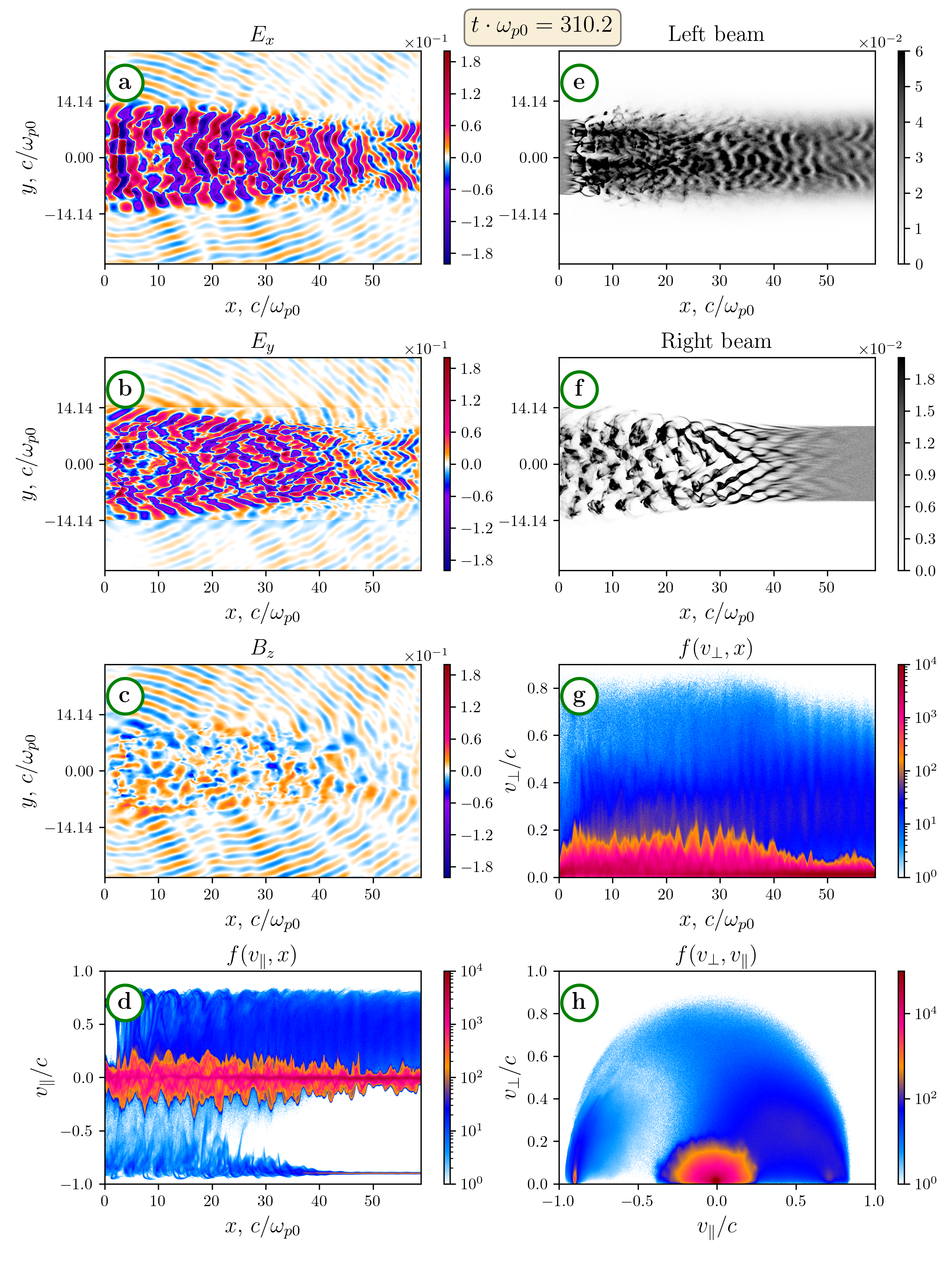}
	\caption{Same as on fig. \ref{fig:pic09}. Relativistic and sub-relativistic beams with thickness $\sigma=6\pi$. Time moment $t\cdot\omega_{p0}\approx310$.}
	
	\label{fig:pic0709}
\end{figure*}

Let us consider how the emission mechanism discussed would work in the case of different colliding beams. For this purpose we carried out a simulation in which the left beam was subrelativistic and the right one was relativistic. The beam sizes $\sigma=6\lambda_\perp$ were equal and the longitudinal length of the system was $L=1500\Delta x\approx59$~$c/\omega_{p0}$. The results obtained are shown in Figure \ref{fig:pic0709}. As in the previous regimes the emission efficiency (Figure \ref{fig:eff0709}) is  a percentage of the beam power, and the spectrum (Figure \ref{fig:fft0709}) is dominated by radiation near the second harmonic of the plasma frequency, as expected.

Also let us note the significant change in the phase portrait of the beams seen in figure \ref{fig:pic0709}d caused by interaction with large amplitude plasma oscillations. In this case, we can not any more consider that the oscillations are excited by beams with a given at the time of injection velocity $v_b$. Instead, it is more correct to consider some effective beam velocity. Let us define this velocity $v_b^{eff}$ as follows. Assume that the injected electron beam after propagating through a region with previously excited plasma oscillations acquires a complex distribution function determined by both the energy loss and heating of the beam and the process of trapping of the beam in the plasma wave. At the considered location, the beam undergoes the two-stream instability and excites plasma waves with a frequency $\omega_b'\approx\omega_{p0}$ and a wave number $k_\parallel'$. By effective beam velocity we mean the velocity of the monoenergetic beam at which $k_\parallel'=\omega_{p0}/v_b^{eff}$. Due to the change in velocity, the wave number of plasma oscillations excited also changes. However, in the case of symmetric counter beams this does not change the angle of radiation generation due to the identical change of $v_b^{eff}$ for both beams.


Let us estimate the emission angle predicted from the theory. The longitudinal wave number of the radiation-generating perturbation is calculated from the wave numbers of the oscillations excited by each beam (table \ref{tab:params}):
\begin{align*}
	k_\parallel&=k_\parallel^{(1)}-k_\parallel^{(2)}\\
	k_\parallel&=1.46-1.14=0.32.
\end{align*}

\begin{figure*}
	\centering
	\includegraphics[width=0.49\linewidth]{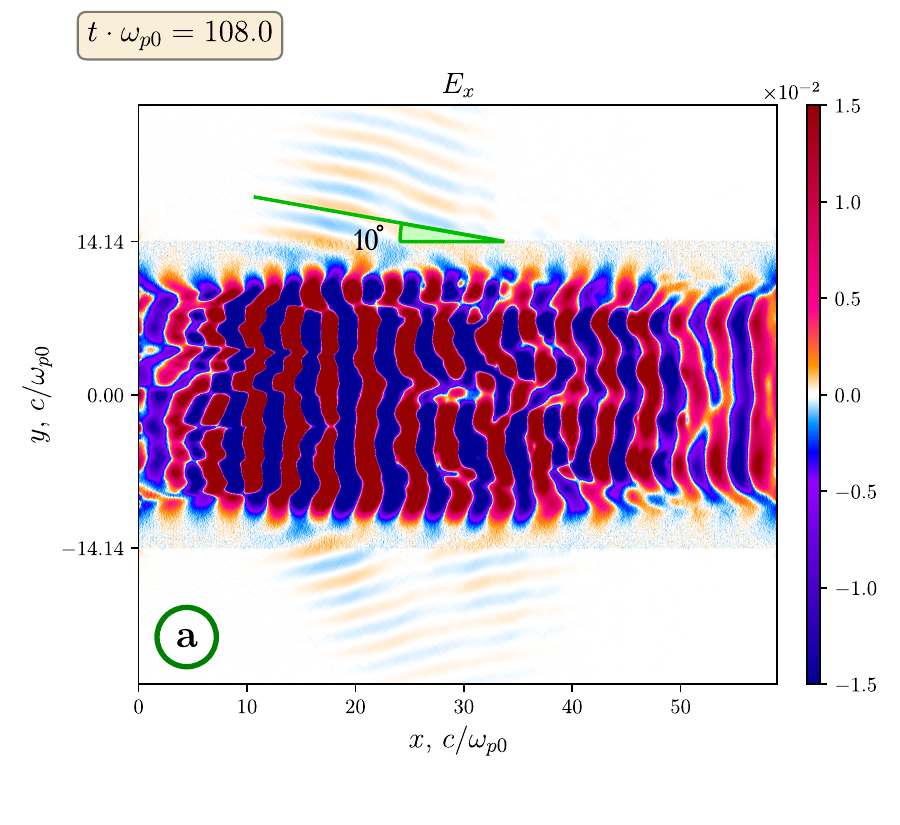}
	\hfill
	\includegraphics[width=0.49\linewidth]{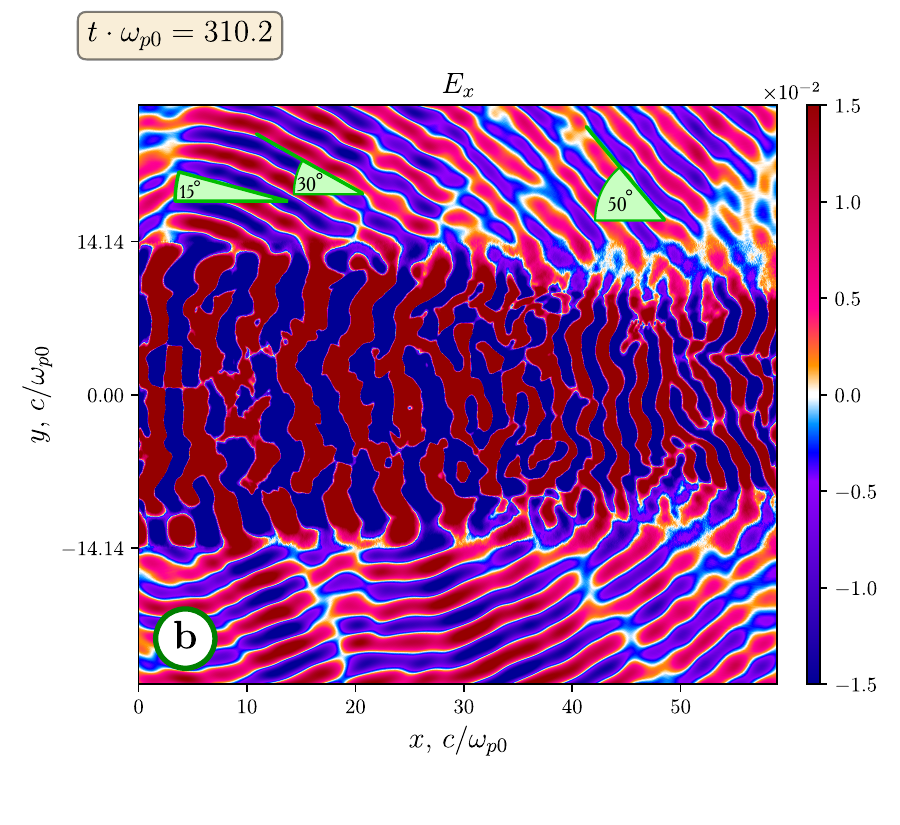}
	\caption{The field $E_x$ for counter-streaming relativistic and sub-relativistic beams. The initial moment of development of the instability and the later one are shown. Angles of radiation with respect to the system's axis are marked in green.}
	\label{fig:pic0709angles}
\end{figure*}
The oscillation frequency can be obtained from the sum of the theoretical frequencies $\omega_r=0.94+0.87=1.81$. Then the transverse wave number of radiation should be $k_\perp=\sqrt{\omega_r^2-k_\parallel^2}=1.78$ and the angle of emission \ref{eq:angle}: $\alpha\approx10$ degrees. Radiation at this angle is indeed observed, but only at the very beginning of the instability development process (Fig. \ref{fig:pic0709angles}a). Then, the distribution of the freshly injected subrelativistic beam changes dramatically under the action of previously excited plasma oscillations (Fig. \ref{fig:pic0709}d). Since, in contrast to the case of symmetric beam injection (section \ref{subsec:counter07}), in this case the second beam (relativistic) does not experience a symmetric change in effective velocity, a significant change in the emission angle is observed (Fig. \ref{fig:pic0709angles}b). The further away the sub-relativistic beam is from the injection site, the lower its effective velocity and the larger the emission angle. Near the region of relativistic beam injection one observes an emission angle of about 50 degrees. Since at this place the effective speed of the right beam has not yet had time to change essentially, from the angle of radiation we can estimate the effective speed of the sub-relativistic beam which was $v_{eff}\approx0.3c$.

Note that this does not mean that the total beam energy has decreased by nearly 90~\% after passing through the plasma, dropping to a value corresponding to the velocity $v\approx0.3c$. Based on measurements of the beam energy flux at the entrance and exit of the system, its energy loss can be estimated at $\approx33$~\%. Some of this energy was lost to radiation and the other part to heating of plasma and excitation of non-emitting plasma oscillations.

\section{Discussion and Conclusion}\label{sub:disc_conc}

Let us briefly summarise the main results of this paper:
\begin{enumerate}
	\item The emission mechanism during the interaction of counter-propagating plasma waves with different transverse structure is effectively realized for plasmas with two counter-streaming electron beams in the regime of oblique small-scale instability dominance.
	\item These regimes can be predicted by the exact linear theory of beam-plasma instability without full-scale PIC simulations. 	
	\item On the scales considered in this article there is no dependence of the generation efficiency on the thickness of electron beams. Also the spectrum of radiation is the same, since the generation takes place at harmonics of plasma frequency starting from the second one. For such radiation the surrounding plasma is transparent.
	\item Beam-to-radiation power conversion efficiency for all modes considered is at the level of a few percent, which is a sufficiently high value for beam-plasma systems.
\end{enumerate}

In work \cite{Annenkov2020} emission at the second harmonic of plasma frequency  was also considered in the system with counter-streaming beams in plasma, but due to the three-wave merging process of plasma waves into electromagnetic ones. It has been demonstrated that achieving radiation generation with an efficiency of a few percent of the beam power requires precise parameter selection for the system. In this regime the most unstable modes of plasma oscillations lie exactly in a certain region of the spectrum for which the necessary three-wave process is allowed. Even an insignificant change of parameters of the system, which leads to leaving this region, means a decrease of efficiency of radiation generation in times. Thus, this mechanism appears to be extremely sensitive to the parameters of the system and even a small change of them can interrupt the generation of radiation. In contrast to it, the mechanism considered in this work does not require exact selection of system parameters and should work with similar efficiency in a wide range of parameters, at which oblique modes of plasma oscillations dominate.

Let us discuss what practical recommendations can be made on the basis of these results. There are two areas, for which the considered mechanism of radiation generation may be relevant. The first one is generation of powerful, frequency-tunable narrow-band radiation in laboratory plasma. Using the exact linear theory, it is possible to choose parameters of the beam-plasma system and, with the help of PIC simulation, to estimate the efficiency of generation and the beam relaxation length, which is necessary for the design of radiation sources.

The second one is the sources of non-thermal radio emission in astrophysical systems and especially in solar plasma. Systems with counter electron fluxes can arise at interaction of the curved front of shock waves with magnetic field lines or in the case of closely spaced regions of intense energy release due to magnetic reconnection. By estimating the parameters of the surrounding plasma, such as thermal and X-ray radiation, from measurements, and utilizing existing models of similar systems, one can incorporate these parameters into an exact linear algorithm. This algorithm calculates the growth rate of the beam-plasma instability and provides an estimate of the possibility of radiation generation by the discussed mechanism.

Of further interest in the research of this generation mechanism is the question about dependence of radiation parameters on transverse length of excited oscillations. Study of regimes with smaller and larger energy beams is also relevant to this investigation, as well as consideration of the relaxation process in plasma with large and small-scale density inhomogeneities and  full-scale 3D3V PIC simulations.

\section{Acknowledgments}
The work was supported by the Foundation for the Advancement of Theoretical Physics and Mathematics ''BASIS''. Simulations were performed using the computing resources of the Center for Scientific IT-services ICT SB RAS.

%

\end{document}